\documentclass[11pt,onecolumn]{scrartcl}
\usepackage{float,afterpage,longtable,amssymb,amsmath,eucal}
\usepackage{graphicx}
\usepackage[latin1]{inputenc}
\usepackage[english]{babel}
\usepackage[usenames]{color}
\usepackage[colorlinks=true,linkcolor=blue]{hyperref}
\usepackage{natbib}
\usepackage{nicefrac}
\usepackage{setspace}
\usepackage{url}
\usepackage{lscape}
\usepackage{lineno}
\usepackage{longtable}
\renewcommand{\theta}{\vartheta}

\newcommand{\Secref}[1]{Section \ref{#1}}
\newcommand{\secref}[1]{section \ref{#1}}
\newcommand{\Tabref}[2][]{Table \ref{#2}#1}

\newcommand{\ome}{\mbox{$\mathcal{O}$}}
\newcommand{\eaa}{\mbox{$\mathcal{A}$}}
\newcommand{\hemi}{\mbox{$\mathcal{H}$}}
\newcommand{\hcmb}{\mbox{$\mathcal{H}_{cmb}$}}
\newcommand{\hsur}{\mbox{$\mathcal{H}_{sur}$}}
\newcommand{\rmeaa}{\mbox{${Rm^\star}$}}
\newcommand{\tp}[1]{\mbox{$\times10^{#1}$}}

\title{A hemispherical dynamo model : Implications for the Martian crustal magnetization}
\author{Wieland Dietrich$^{1,2}$ \& Johannes Wicht$^{2}$ \vspace{0.2cm}
 \\ {\small $ ^2$Max-Planck-Institut f\"{u}r Sonnensystemforschung, }
 \\ {\small Max-Planck-Strasse 2, 37191 Katlenburg-Lindau, Germany}
 \\ {\small $ ^1$Institute for Geophysics,}
 \\ {\small  University of G\"{o}ttingen, 37073 G\"{o}ttingen, Germany}}
\begin{document}
\maketitle
\begin{abstract}
Mars Global Surveyor measurements revealed that the Martian crust is strongly
magnetized in the southern hemisphere while the northern hemisphere
is virtually void of magnetization.
Two possible reasons have been suggested for this dichotomy:
A once more or less homogeneously magnetization may have been destroyed
in the northern hemisphere by, for example, resurfacing or impacts.
The alternative theory we further explore here assumes that the dynamo itself
produced a hemispherical field \citep{Stanley2008,Amit2011}.
We use numerical dynamo simulations to study under which conditions
a spatial variation of the heat flux through the core-mantle boundary (CMB)
may yield a strongly hemispherical surface field. We assume that the early
Martian dynamo was exclusively driven by secular cooling and we mostly concentrate
on a cosine CMB heat flux pattern with a minimum at the north pole, possibly
caused by the impacts responsible for the northern lowlands.
This pattern consistently triggers a convective mode
which is dominated by equatorially anti-symmetric and axisymmetric (EAA, \citet{Landeau2011}) thermal winds. Convective up- and down-wellings and thus radial magnetic field production then tend to concentrate in the southern hemisphere which is still cooled efficiently
while the northern hemisphere remains hot. The dynamo changes from an $\alpha^2$- for a homogeneous CMB heat flux to an $\alpha\Omega$-type in the hemispherical configuration. These dynamos reverse on time scales of about 10 kyrs. This too fast to allow for the more or less unidirectional magnetization of thick crustal layer required to explain the strong magnetization in the southern hemisphere.

\end{abstract}
 
%\begin{keyword}
%Ancient Martian Dynamo \sep Hemispherical magnetic field \sep CMB heat flux anomaly
%\end{keyword}
\maketitle

\begin{linenumbers}

\section{Introduction}
Starting in 1998 the space probe Mars Global Surveyor (MGS) delivered vector
magnetic field data from orbits between $185$ and $400$ km above the planets surface
\citep{Acuna1999}. The measurements reveal a strong but heterogeneous
crustal magnetization \citep{Acuna1999,Connerney2001}. The more
strongly magnetized rocks are mainly localized in the southern hemisphere
where the crust is thick and old. The northern hemisphere is covered by a
younger and thinner crust which is much weaker magnetized.

Two alternative types of scenarios are discussed to explain this dichotomy.
One type explores the possibility that an originally more or less
homogeneous magnetization was partly destroyed by resurfacing events after
the demise of the internal dynamo. Based on the fact that the
Hellas and Argyre impact basins are largely void of magnetization,
\citet{Acuna2001} conclude that the dynamo stopped operating in the
early Noachian before the related impact events happened roughly $3.7-4$ Gyrs
ago. Volcanic activity and crustal spreading are two other
possibilities to explain the lack of strong magnetization in certain
surface areas \citep{Lillis2008b, Mohit2004}, in particular the
northern hemisphere after the dynamo cessation.

The alternative scenario explains the dichotomy by an ancient Martian dynamo
that inherently produced a hemispherical magnetic field.
Numerical dynamo simulations by \citet{Stanley2008} and \citet{Amit2011}
show that this may happen when more heat is allowed
to escape the core through the southern than through the northern
core mantle boundary (CMB). Such north/south asymmetry can for example be caused by larger impacts or low-degree mantle convection \citep{Roberts2006,Keller2009,Yoshida2006}. Due to depth-dependent viscosity and a possible endothermic phase transition \citep{Harder1996} Martian mantle convection may be ruled in an extreme case by one gigantic plume typically
evoked to explain the dominance of the volcanic Tharsis region. However, the single plume convection might have developed after the dynamo ceased.
Due to the hotter temperature of the rising material the CMB heat flux
can be significantly reduced under such a plume.
Though Tharsis is roughly located in the equatorial region it could
nevertheless lead to magnetic field with the observed north-south symmetry, as
we will show in the following. 

The possible effects of large impacts on planets and the dynamo in
particular are little understood.
\citet{Roberts2009} argue that impacts locally heat the underlying mantle
and thereby lead to variations in the CMB heat flux. 
Large impacts may also cause a demise of the dynamo by reducing the CMB heat
flux below the value where subcritical dynamo action is still possible
\citep{Roberts2009}. The deposition of heat in the outer parts of the core
by impact shock waves could lead to a stably stratified core and thereby also
stop dynamo action \citep{Arkani-Hamed2010} for millions of years until the
heat has diffused out of the core. If the iron content of the
impactor is large enough it may even trigger a dynamo \citep{Reese2010}.

The thermal state of the ancient Martian core is rather unconstrained \citep{Breuer2010}. Analysis of Martian meteorites suggests a significant sulphur content and thus a high core melting temperature \citep{Dreibus1985}. Mars may therefore never have grown a solid inner core, an assumption we also adopt here \citep{Schubert1990, Breuer2010}. The ancient Martian dynamo was
then exclusively driven by secular cooling and radiogenic heating and
has stopped operating when the CMB heat flux became
subadiabatic \citep{Stevenson1983}. Run-away solidification or light element saturation may explain the dynamo cessation in the presence an inner core.

The geodynamo, on the other hand, is predominantly driven
by the latent heat and light elements emanating from the growing inner
core front. Secular cooling and
radiogenic heating is typically modeled by homogeneously distributed
internal buoyancy sources while the driving associated to inner core growth
is modeled by bottom sources \citep{Kutzner2000}.
These latter sources have a higher Carnot efficiency and would
likely have kept the Martian dynamo alive if the planet would have
formed an inner core.

Several authors have explored the influence of a CMB heat flux
pattern on dynamos geared to model Earth and report that they
can cause hemispherical variations in the secular variation
\citep{Bloxham2000,Christensen2003,Amit2006},
influence the reversal behavior \citep{Glatzmaier1999,Kutzner2004}
or lead to inhomogeneous inner core growth \citep{Aubert2008}.
However, the effects where never as drastic as those reported by
\citet{Stanley2008} or \citet{Amit2011}. In the work of
\citet{Amit2011} the reason likely is the
increased susceptibility of internally driven dynamos to
the thermal CMB boundary condition \citep{Hori2010}.
\citet{Stanley2008} retain bottom driving and employed a particularly
strong heat flux variation to enforce a hemispherical field

Here, we follow \citet{Amit2011} in exploring the effects of
a simple sinusoidal CMB heat flux variation on a dynamo model
driven by internal heat sources. The main scope of this paper is
to understand the particular dynamo mechanism, to explore its
time dependence, and to extrapolate the results to the Martian
situation. \Secref{Model} introduces our model, whereas \secref{hemisphericalaction}
describes the effects of the CMB heat flux anomaly on the convection and 
the induction process. In \secref{applymars} we explore the applicability to the
ancient Martian dynamo. The paper closes with a discussion in \secref{Discussion}.

\section{Numerical Model}
\label{Model}
Using the MagIC code \citep{Wicht2002,Christensen2007}, we model the Martian core as a viscous, electrically
conducting and incompressible fluid contained in a rotating spherical shell with inner core radius $r_{icb}$ and outer
radius $r_{cmb}$. Conservation of momentum is described by the dimensionless Navier-Stokes equation
for a Boussinesq fluid:
\begin{equation}
E \left( \frac{\partial \vec u}{\partial t} + \vec u \cdot \vec \nabla \vec u \right)
= -\vec \nabla \Pi + E \nabla^2 \vec u - 2 \hat z \times \vec u + \frac{Ra E}{Pr} \frac {\vec r}{r_{cmb}} T
+ \frac 1 {Pm} (\vec \nabla \times \vec B ) \times \vec B \,
\label{nseq}
\end{equation}
where $\vec u$ is the velocity field, $\Pi$ the generalized pressure, $\hat z$
the direction of the rotation axis, $T$ the super-adiabatic temperature and
$\vec B$ the magnetic field.

The conservation of energy is given by
\begin{equation}
\label{heat}
\frac {\partial T}{\partial t} + \vec u \cdot \vec \nabla T= \frac 1 {Pr} \nabla^2 T
+ \epsilon \\,
\end{equation}
where $\epsilon$ is a uniform heat source density.
The conservation of magnetic field is given by the induction equation
\begin{equation}
\frac {\partial \vec B}{\partial t} = \vec \nabla \times \left( \vec u \times \vec B \right)
+ \frac 1 {Pm} \nabla^2 \vec B \ .
\label{ind_eq}
\end{equation}

We use the shell thickness $D=r_{cmb}-r_{icb}$ as  length scale, the
viscous diffusion time $D^2/ \nu$ as time scale and $(\rho \mu \lambda \Omega)^{1/2}$
as the magnetic scale. The mean superadiabatic CMB heat flux density $q_0$ serves
to define the temperature scale $q_0 D / c_p\rho\kappa$.
Here, $\nu$ is the viscous diffusivity,
$\rho$ the constant background density, $\mu$ the magnetic permeability, $\lambda$ the
magnetic diffusivity, $\Omega$ the rotation rate, $\kappa$ the thermal diffusivity
and $c_p$ the heat capacity.

Three dimensionless parameters appear in the above system: the Ekman number
$E=\nu / \Omega D^2 $ is a measure for the relative importance of viscous versus Coriolis
forces while the flux based Rayleigh number $Ra=\alpha g_0 |q_0| D^4 / \rho c_p\kappa^2\nu$ is a measure
for the importance of buoyancy. The Prandtl number $P=\nu / \kappa$ and the
magnetic Prandtl number  $Pm=\nu / \lambda$  are diffusivity ratios.

An inner core with $r_{icb}/r_{cmb}=0.35$ is retained for numerical reasons \citep{Hori2010}, but to minimize its influence the heat flux from the inner core is set to zero. The secular cooling and radiogenic driving is modeled by the
homogeneous heat sources $\epsilon$ appearing in
\ref{heat} \citep{Kutzner2000}. Furthermore we assume an electrically insulating inner core to avoid an additional sink for the magnetic field.
We use no-slip, impermeable flow boundary conditions and match $\vec B$ to a potential
field at the outer and inner boundary. 
The results by \citet{Hori2010} and \citet{Aubert2009} suggest
that this is a fair approximation to model a dynamo without inner core since
an additional reduction of the inner core radius has only a minor impact.
The effective heat source $\epsilon$ is
chosen to balance the mean heat flux $q_0$ through the outer boundary:
\begin{align}
4 \pi^2 r_{cmb}^2\;q_0 &=-Pr \frac 4 3 \pi (r_{cmb}^3-r_{icb}^3)\;\epsilon \ .
\end{align}
Note that $q_0$ is generally negative. The CMB heat flux pattern is modeled in terms of spherical harmonic contributions with amplitude $q_{lm}$, where $l$ is the degree and $m$ the spherical harmonic order. Here we mostly concentrate on a variation along colatitude $\theta$ of the form $q_{10} \cos{\theta}$
with negative $q_{10}$ so that the minimum (maximum) heat flux is located
at the north (south) pole. This is the most simple pattern to break the
north/south symmetry and has first been used by \citet{Stanley2008} in the
context of Mars. We also explore the equatorially
symmetric disturbance $q_{11} \sin{\theta} \sin{\phi}$, which breaks the east/west
symmetry, and a superposition of $q_{10}$ and $q_{11}$ to describe
a cosine disturbance with arbitrary tilt angle
\begin{align}
\alpha = \operatorname{arctan} (|q_{11}| / |q_{10}|)\;\;.
\label{defalpha}
\end{align}
In the following we will characterize the amplitude of any disturbance by
its maximum relative variation amplitude in percent
\begin{align}
g = 100 \% \max(|\delta q|) / |q_0|\;\;.
\end{align}
We vary $g$ up to $300\%$, the value used in \citet{Stanley2008}.
For variations beyond $100\%$ the heat flux becomes
subadiabatic in the vicinity of the lowest flux. For severely subadiabatic
cases this may pose a problem since dynamo codes typically solve for
small disturbances around an adiabatic background state \citep{Braginsky1995}.
The possible implication of this have not been explored so far and
we simply assume that the model is still valid. Since the main effects
described below do not rely on $g>100\%$ this is not really an issue here.

The hemispherical mode triggered by the heat flux variation
is dominated by equatorially anti-symmetric and axisymmetric thermal winds \citep{Landeau2011}. Classical columnar convection found for a homogeneous heat flux, on the other hand,
is predominantly equatorial symmetric and non-axisymmetric, at least at
lower Rayleigh numbers. We thus use the relative
equatorial anti-symmetric and axisymmetric (EAA) kinetic energy to identify the hemispherical mode:
\begin{align}
\eaa = \frac { \sum_{l_{odd}, m=0} E_{lm} } { \sum_{lm} E_{lm} }  \ ,
\label{defeaa}
\end{align}
where $E_{lm}$ is the rms kinetic energy carried by a flow mode of
spherical harmonic degree $l$ and order $m$.

For a homogeneous outer boundary heat flux the dynamo is to first order of an $\alpha^2$-type 
where poloidal and toroidal fields are produced in the individual convective columns \citep{Olson1999}. As the hemispherical flow mode takes over, the
$\Omega$-effect representing the induction of axisymmetric toroidal magnetic field via axisymmetric shearing becomes increasingly important. We measure its relative contribution to
toroidal field production by
\begin{equation}
\ome =   \frac { \left[(\bar{\vec{B}}\cdot \vec \nabla )\;
\bar{u}_\phi \right]_{\mbox{\scriptsize tor}}^{\mbox{\scriptsize rms}}}
{ \left[(\vec B\cdot \vec \nabla ) \vec u \right]^{\mbox{\scriptsize rms}}_{\mbox{\scriptsize tor}}}  \ .
\label{defomega}
\end{equation}
The lower index tor and upper index rms indicate that rms values of the
toroidal field production in the shell are considered.

For quantifying to which degree the Martian crustal magnetization and the
poloidal magnetic fields in our dynamo simulations are concentrated in
one hemisphere we use the hemisphericity measure
\begin{equation}
\hemi(r)= \left\vert \frac {  B_r ^{N}(r) - B_r ^{S}(r)  }
{  B_r ^{N}(r) +  B_r ^{S}(r)  } \right\vert \ ,
\label{Hdef}
\end{equation}
where $B_r ^{N}(r)$ and $B_r ^{S}(r)$ are the surface integral over the
unsigned radial magnetic flux in the northern and southern hemispheres,
respectively.
According to this definition both a purely equatorially symmetric and
a purely equatorially anti-symmetric field yield $\mathcal{H}=0$.
For $\mathcal{H}=1$ the flux is strictly concentrated in one hemisphere
which requires a suitable combination of equatorially symmetric and
anti-symmetric modes \citep{Grote2000}. A potential field extrapolation is used to
calculate $\mathcal{H}$ for radii above $r_{cmb}$, for example the surface hemisphericity \hsur. 

\Tabref{Tab1} provides an overview of the different parameter combinations
explored in this study along with $\mathcal{A}$, $\mathcal{O}$,
\hcmb, \hsur, the Elsasser number
$\Lambda=B^2 / \mu_0 \lambda \rho \Omega$ and the field strength at the Martian surface in nano Tesla $\bar{B}_{sur}$. 
Column $14$ lists the respective (if present) dimensionless oscillation frequencies given in units of magnetic diffusion time.

We mostly focus on simulations at $E=10^{-4}$ where the relatively moderate
numerical resolution still allows to extensively explore the
other parameters in the system. A few cases at $E=3\times10^{-5}$ and
$E=10^{-5}$ provide a first idea of the Ekman number dependence.The last line in table \ref{Tab1} gives estimates for the Rayleigh, Ekman and magnetic Prandtl number of Mars, based on the (rather uncertain) properties of Mars \citep{Morschhauser2011}.

{\footnotesize
\begin{longtable}{ccc|ccccccccccc}
$E$ & $Ra$ & $Pm$ & $g$ & $\alpha$ & $Rm$ &\rmeaa & $\Lambda$ &$\mathcal{A}$& $\mathcal{O}$ & $\bar{B}_{sur}$ & $\mathcal{H}_{sur}$&$\mathcal{H}_{cmb}$ &  freq.  \\
\hline
1e-4   & 7e6    &  2  &  0    &  0   &  54.6   &  0.24  &  -     &  2.24e-5  &  -      &  -       &  -     & -       & -     
\\
       &        &     &  100  &  0   &  133.5  &  122.6  &  0.1   &  0.85     &  0.32   &  803.2   &  0.1   & 0.21   & -     \\
       &        &  5  &  0    &  0   &  117.1  &  3.95  &  9.79  &  1.93e-3  &  0.21   &  62510   &  4e-4  & 3e-3    & -     \\
       &        &     &  100  &  0   &  326.9  &  301.5  &  0.97  &  0.85     &  0.66   &  1469    &  0.1   & 0.35   & ?     \\
       &        &     &  200  &  0   &  449.5  &  417.5  &    -   &  0.84     &    -    &    -     &    -   &  -     & -     \\
       &        &     &  100  &  90  &  230.8  &  -     &  2.22  &  5e-3     &  0.20   &  7264    & -      &  -     & 18.84 \\  
\hline
       & 2.1e7  &  2  &  0    &  0   &  105.9  &  6.42  &  4.95  &  3.64e-3  &  0.24   &  64635   &  1.0e-3& 0.03    & -     \\
       &        &     &  60   &  0   &  178.1  & 149.8  &  6.24  &  0.72     &  0.63   &  14017   &  0.12  & 0.38    & -     \\
       &        &     &  80   &  0   &  228.3  & 189.7  &  1.06  &  0.74     &  0.53   &  1684    &  0.17  & 0.61   & 10.69 \\
       &        &     &  100  &  0   &  247.6  & 213.6  &  0.15  &  0.73     &  0.58   &  1001    &  0.21  & 0.55   & ?     \\
       &        &     &  200  &  0   &  313.3  & 272    &  0.19  &  0.76     &  0.77   &  689     &  0.79  & 0.8    & 56.27 \\
       &        &     &  100  &  90  &  160.2  &    -    &  2.64  &  6.6e-3   &  0.18   &  699     & -      &  -   & ?     \\
\hline
       &   4e7  &  1  &  100  &  0   &  169.3  &  143.1 &  0.2   &  0.73     &  0.53   &  922     &  0.21  & 0.22    & -     \\
       &        &     &  200  &  0   &  206.5  &  171.5 &   -    &   0.7     &   -     &   -      &   -    &  -    &  ?     \\
       &        &  2  &  0    &  0   &  155.6  & 3.58   &  6.26  &  2e-3     &  0.18   &  58349   &  3.0e-3& 0.05    & -     \\
       &        &     &  60   &  0   &  283.4  & 252.4  &  4.18  &  0.59     &  0.65   &  6154    &  0.26  & 0.6    & 13.96 \\
       &        &     &  100  &  0   &  338.1  & 307.2  &  2.64  &  0.78     &  0.76   &  2219    &  0.52  & 0.77   & 40.83 \\
       &        &     &  200  &  0   &  409.8  & 350    &  1.16  &  0.74     &  0.75   &  1628    &  0.74  & 0.75   & 62.6  \\
       &        &     &  100  &  90  &  217.3  &   -    &  1.47  &  8.4e-3   &  0.21   &  10071   & -      &  -     & 20.77 \\
       &        &     &  200  &  90  &  226.5  &   -    &  5.41  &  4.1e-3   &  0.24   &  10934   & -      &  -      & -     \\
       &        &  5  &  100  &  0   &  837.5  &  749.5 &  6.83  &  0.81     &  0.8    &  3493    &  0.7   &  0.65  & 79.87 \\
\hline
       &  8e7   &  2  &  0    &  0   &  228.7  & 6.4    &  7.06  &  9e-3     &  0.18   &  60036   &  3.3e-3& 0.07    & -     \\
       &        &     &  60   &  0   &  400.2  & 343.6  &  5.5   &  0.74     &  0.65   &  9268    &  0.41  & 0.72   & 26.97 \\
       &        &     &  100  &  0   &  457.6  & 403.2 &  2.97  &  0.79     &  0.73   &  3240    &  0.68  & 0.73    & 61.3  \\
       &        &     &  100  &  90  &  297.5  &   -    &  3.73  &  6.4e-3   &  0.20   &  16424   & -      &  -     & 24.5  \\
\hline
       &  2e8   &  1  &  0    &  0   &  251.8  &  54.4  &  -    &  0.05     &  -      &    0     &   -    &  -      & -     \\
       &        &     &  60   &  0   &  309.9  &  230.1 &  2.14  &  0.63     &  0.56   &  6095    &  0.15  & 0.56   & -     \\
       &        &     &  100  &  0   &  343.5  &  276.3 &  0.34  &  0.64     &  0.65   &  1119    &  0.42  & 0.72   & 24.32 \\
       &        &     &  100  &  90  &  270.4  &   -    &  0.77  &  4e-3     &  0.21   &  7954    & -      &  -     & 17.95 \\
\hline
3e-5   &  1e8   &  2  &  0    &  0   &  137.1  &  2.88  &  7.68  &  5.7e-3   &  0.19   & 80508    &  1e-3  & 0.03   & -     \\
       &        &     &  60   &  0   &  210.7  &  48.2  & 12.31  &  0.5      &  0.53   & 25567    &  0.1   & 0.23   & -     \\
       &        &     &  100  &  0   &  360.4  &  324.2 &  5.07  &  0.81     &  0.67   &  5157    &  0.12  &  0.46  & 10.34 \\
       &        &     &  100  &  90  &  199.5  &    -   &  1.4   &  3.1e-3   &  0.16   & 10657    & -      &  -     & ?     \\
\hline
3e-5   &  4e8   &  2  &  0    &  0   &  316.9  &  6.62  &  12.2  &  5.5e-3   &  0.26   & 71085    &2.1e-3  & 0.07 & -     \\
       &        &     &  60   &  0   &  517.1  &  401.8 & 29.9   &  0.64     &  0.49   & 30528    &  0.24  & 0.51   & -     \\
       &        &     &  100  &  0   &  769.1  &  682   &  6.04  &  0.76     &  0.70   & 4584     &  0.62  & 0.76   & 87.6  \\
\hline
1e-5   &  4e8   &  2  &  0    &  0   &  234.9  & 586  &  13.58  &  0.01   &  0.18   & 88763    &6e-3  & 0.09   & -     \\
       &        &     &  50   &  0   &  292.5  &  146.2 & 19.07   &  0.18     &  0.23   & 69618    &  0.03  & 0.22 & -     \\
       &        &     &  100  &  0   &  441.1  &  376.4   & 41.7  &  0.41     &  0.26   & 42194     &  0.07  & 0.30 & -  \\
\hline
 Mars  &        &     &       &      &         &        &        &           &         &          &        &       &       \\
 3e-15 &  2e28  & 1e-6&  ?    &  ?   & 500?    &   ?    &  ?     &  ?        &   ?     & 5000     &  0.45  &  ?    &  ?      \\
%\end{tabular}
\caption{Selection of runs performed. Rm - magnetic Reynolds number, $\Lambda$ - Elsasser number of rms field in full core shell, EAA - relative equatorially antisymmetric and axisymmetric kinetic energy, $\omega^\ast$ - relative induction of toroidal field by shearing,$|B|_{sur}$ - time averaged field intensity at the Martian surface, $\mathcal{H}_{sur}$ and $\mathcal{H}_{cmb}$ - hemisphericity at the surface and CMB, freq. - rough frequency ($2\pi Pm / \tau_{vis}$) if present. Decaying solutions are marked with `-' in the Elsasser number, stationary dynamos with `-' in the frequency. If not a single frequency could be extracted `?' is used.}
\label{Tab1}

\end{longtable}}
%\end{landscape}

\section{Hemispherical Solution}
\label{hemisphericalaction}
We start with discussing the emerging hemispherical dynamo mode promoted
by the $l=1,m=0$ heat flux pattern with minimal (maximal) heat flux at the
north (south) pole concentrating on cases at $E=10^{-4}$, $Ra=4.0 \times 10^{7}$ and
$Pm=2$. The study of \citet{Landeau2011} reports the emergence of the equatorially anti-symmetric and axisymmetric convective mode
if the Rayleigh number is sufficiently high. Note, that the authors used a homogeneous heat flux condition
at the outer boundary. There the amplitude of the hemispherical convection becomes of equal strength compared
to the columnar type in the pure hydrodynamic case and is even more dominant if the magnetic field can act on the flow
\citep{Landeau2011}. 

\subsection{Hemispherical Convection}
\begin{figure}[ht]
\vspace*{2mm}
\par
\begin{center}
\includegraphics[width=0.8\textwidth]{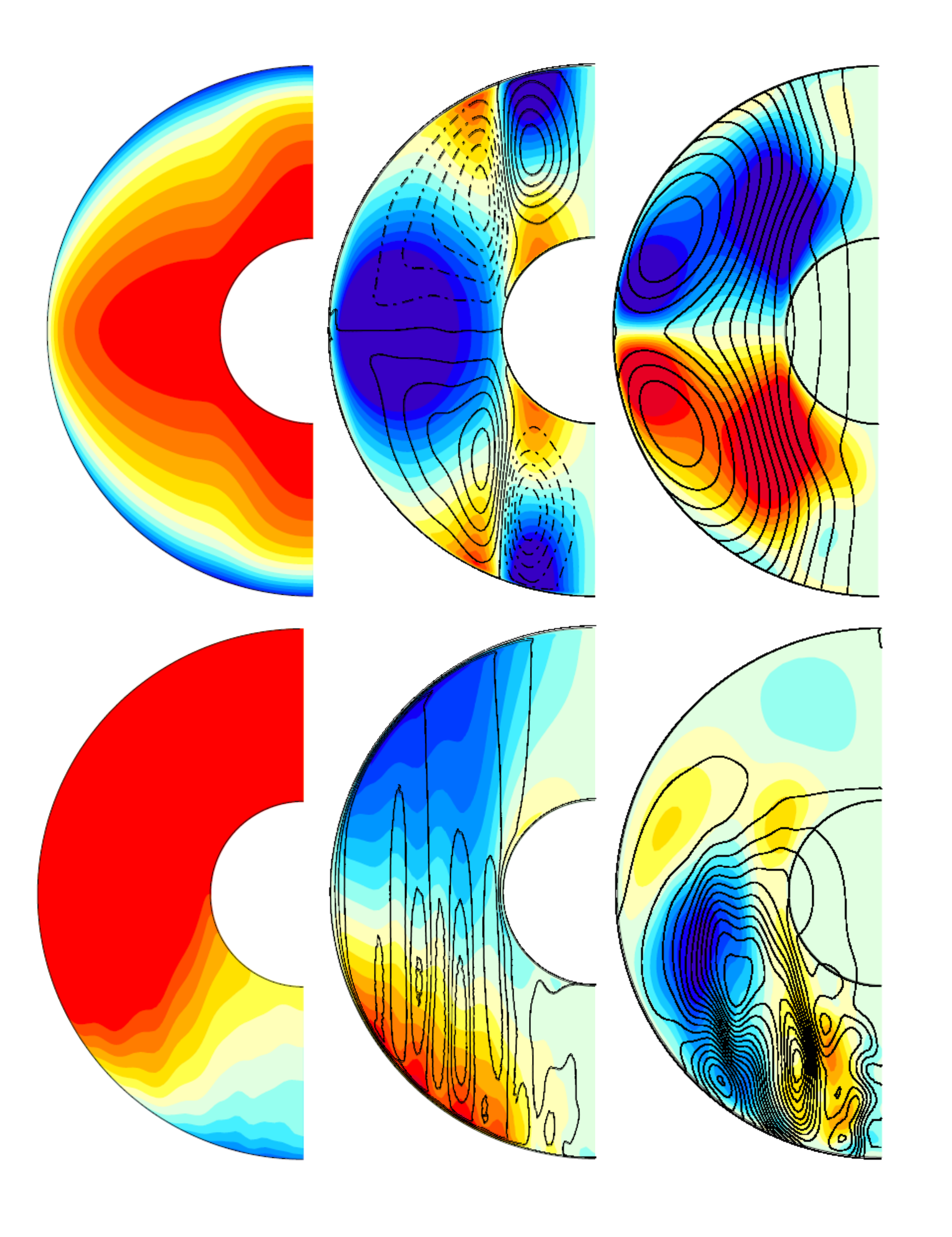}
\end{center}
\caption{Zonal average of the temperature (left plots), zonal flow with
meridional circulation contours (middle plots) and toroidal field with
poloidal field line contours (right plots) for columnar convection
dominated and magnetic dipolar reference case (left) and a typical hemispherical
dynamo solution with the strong EAA symmetry in the flow (right). See the online-version of the article for the color figure.}
\label{zonal}
\end{figure}

\begin{figure}[ht]
\vspace*{2mm}
\par
\begin{center}
\includegraphics[width=1.0\textwidth]{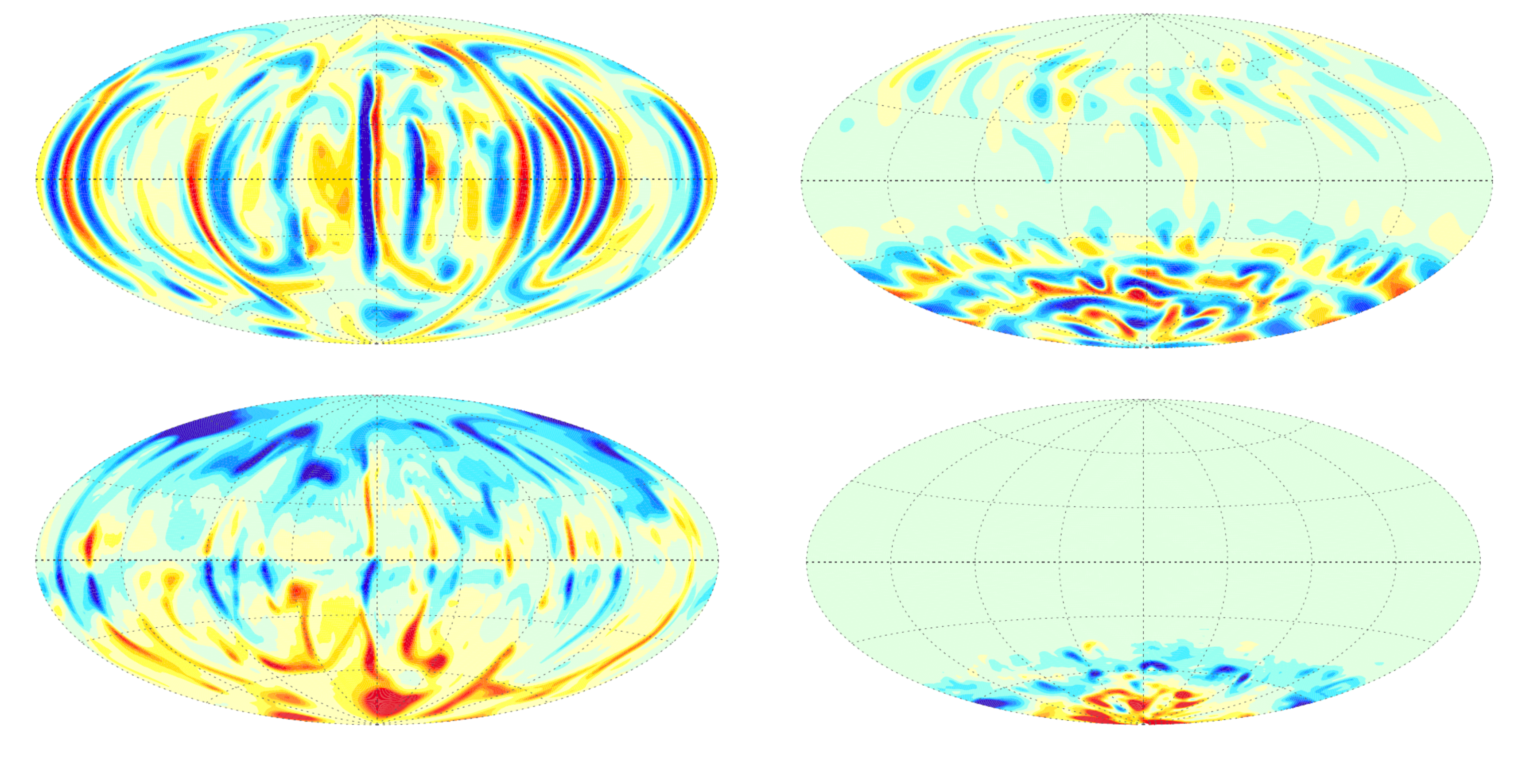}
\end{center}
\caption{Radial flow (top row) at mid-depth and radial field at CMB
(lower row) for the columnar reference case (left) and the hemispherical
dynamo (right), indicates the reduction of the magnetic signature at
the CMB if the radial motions are limited to the southern polar cusp of
high heat flux. Here an Aitoff projection of the spherical CMB is used. See the online-version of the article for the color figure. }
\label{doublehammer}
\end{figure}

Figure \ref{zonal} and \ref{doublehammer} illustrate the typical hemispherical
dynamo configuration emerging at $g=100\%$ and compares this
with the typical dipole dominated dynamo found at $g=0\%$. 
While the southern hemisphere is still cooled efficiently the northern
hemisphere remains hot since radial upwellings and the associated
convective cooling are predominantly concentrated in the southern hemisphere
(figure \ref{doublehammer}, top row). 
The flow pattern changes from classical columnar
solutions to a thermal wind dominated flow which is a direct consequence of the
strong north/south temperature gradient (figure \ref{zonal}, left bottom).
When neglecting inertial, viscous and Lorentz force contributions the
azimuthal component of the curl of the Navier-Stokes equation \eqref{nseq} yields:
\begin{equation}
2 \frac{\partial \bar{u}_\phi}{\partial z} =
\frac {Ra E}{Pr}  \frac{1}{r_{cmb}} \frac{\partial \bar{T}}{\partial \theta}\;\;.
\end{equation}
This is the thermal wind equation and the respective zonal flows will
dominate the solution, indicated by large $\mathcal{A}$ values when the latitudinal temperature gradient is large
enough \citep{Landeau2011}. Since radial flows mainly exist in the southern hemisphere the
production of poloidal and thus radial magnetic field is also concentrated
there. This results in a very hemispherical magnetic field pattern at
the top of the dynamo region (figure \ref{doublehammer}, bottom row).

\begin{figure}[ht]
\vspace*{2mm}
\par
\begin{center}
\includegraphics[width=1.0\textwidth]{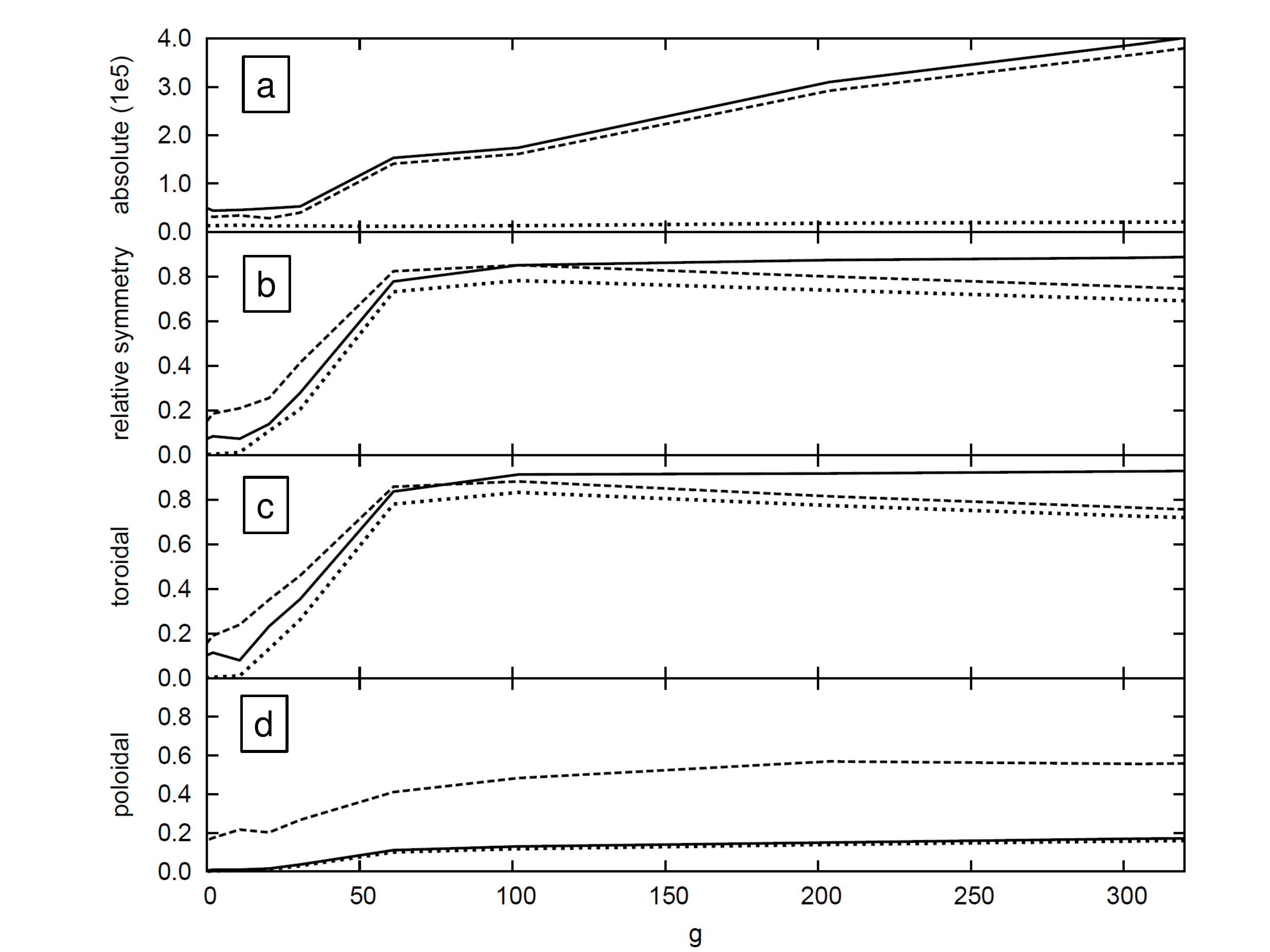}
\end{center}
\caption{Symmetries and amplitude of total kinetic energy and
toroidal/poloidal contributions as a function of $g$.
a) total (solid line),
toroidal (dashed) and poloidal (dotted) kinetic energy; b) relative amount of axisymmetry (solid), equatorial
anti-symmetry (dashed) and the combined symmetries (EAA, dotted)
of the full kinetic energy; c) and d) show the same
but separated into toroidal and poloidal
contributions.}
\label{sym_rel}
\end{figure}

The figure \ref{sym_rel} demonstrates that the toroidal energy
rises quickly with the variation amplitude $g$ while the poloidal energy
is much less effected. The growth of the toroidal energy is explained by
the increasing thermal wind, which is an equatorial anti-symmetric
and axisymmetric (EAA) toroidal flow contribution.
At a disturbance amplitude of $g=60\%$ the EAA contribution accounts for
already $50\%$ of the total kinetic energy (figure \ref{sym_rel}.b ).
The maximum EAA contribution of $\mathcal{A}\approx0.8$ is reached at
$g=100\%$. When further increasing the variation amplitude,
the thermal wind still gains in speed. However, the relative
importance of the EAA mode decreases because the strongest latitudinal
temperature gradient and thus the thermal wind structure moves further
south. This trend is already observed in figure \ref{zonal}.

The equatorial anti-symmetry of the poloidal kinetic energy rises from $10\%$ for $g=0$ to about $50\%$
for $g=100$\% reflecting that upwellings are increasingly concentrated
in one (southern) hemisphere.
The meridional circulation remains weak (figure \ref{sym_rel}.d),
and its contribution to the total EAA energy is minor.

\label{eaadynamics}

\subsection{Dynamo mechanism}

\begin{figure}[ht]
\vspace*{2mm}
\par
\begin{center}
\includegraphics[width=0.8\textwidth]{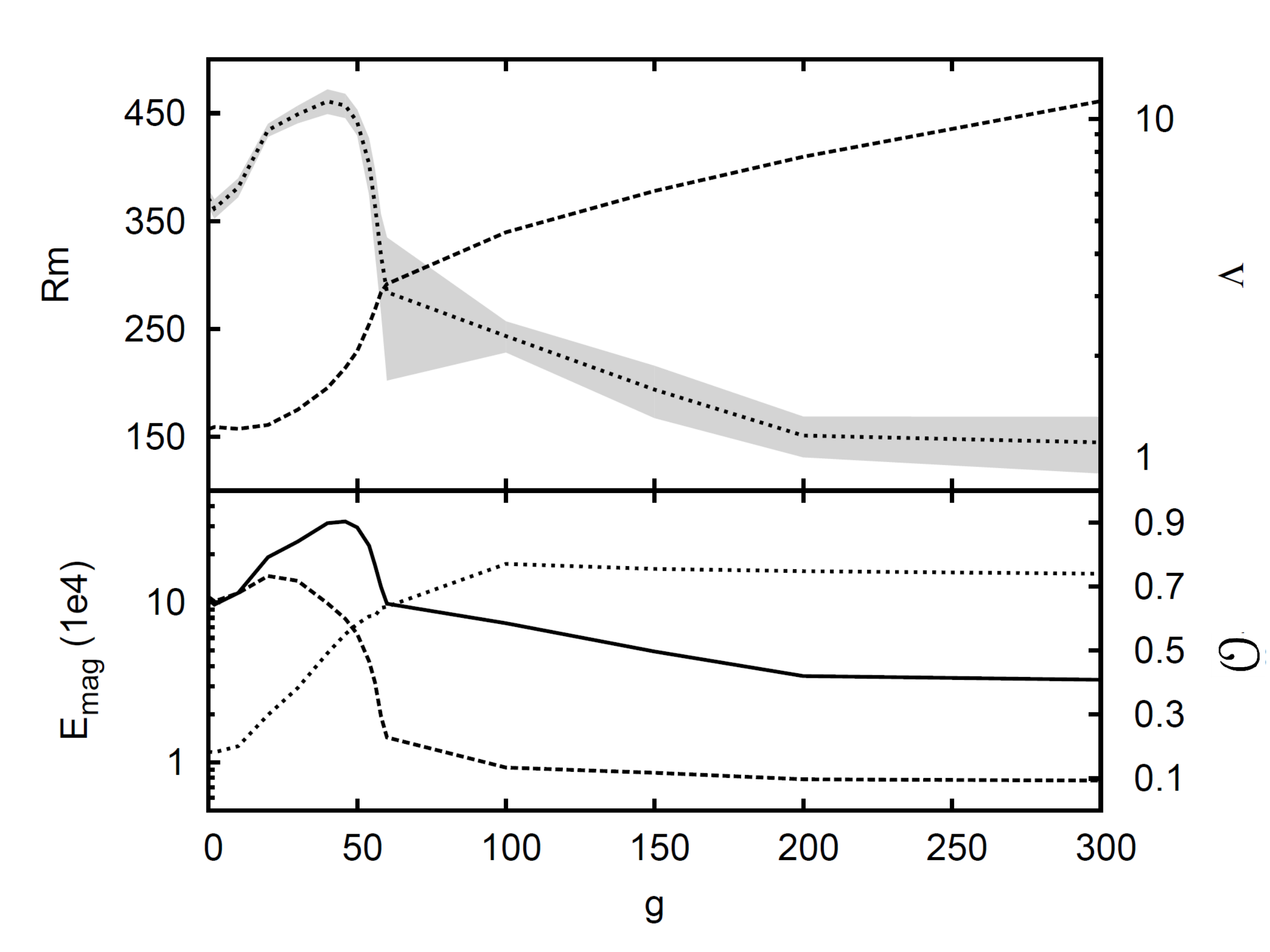}
\end{center}
\caption{Upper panel: Flow amplitude in terms of the magnetic Reynolds number (solid line)
and magnetic field strength in terms of Elsasser number (dashed)) as function
of the CMB heat flux anomaly amplitude $g$, shows the difference between both
dynamo regimes in the efficiency of inducing a dynamo. The hemispherical solution,
with the $\alpha \Omega$-induction contains large amounts of axisymmetric
zonal flows created by the Coriolis force, therefore the kinetic energy is
drastically larger than in the columnar regime ($g=0$). The magnetic energy
decreases, the more the $g$ increases. The gray shade correspond to the standard
deviation due to time variability. \newline
Lower panel: Toroidal (dashed) and poloidal (solid) magnetic field in nondimensional
units and the relative $\Omega$-effect in terms of $\ome$ (dotted) as a
function of $g$ demonstrates the transformation of induction characteristic
from an $\alpha^2$-dynamo at $g=0$ (columnar dynamo) towards
an $\alpha \Omega$-type from $g=60\%$ (hemispherical solution).}
\label{tor_pol_ome}
\end{figure}

The upper panel in figure \ref{tor_pol_ome} demonstrates that the rise in the magnetic Reynolds
number $Rm$, that goes along with the increasing toroidal flow amplitude,
does not necessarily lead to higher Elsasser numbers.
Once more, cases at $E=10^{-4}$, $Ra=4.0 \times 10^{7}$ and $Pm=2$ are depicted here.
For small variation amplitudes up to $g=30$\%  $\Lambda$ still
increases due to the additional $\Omega$-effect associated
to the growing thermal winds. Figure \ref{tor_pol_ome} (lower panel) shows that
the relative contribution of the $\Omega$-effect to toroidal field
production $\mathcal{O}$ grows with $g$.  For $g=0$ it is rather
weak so that the dynamo can be classified as $\alpha^2$ \citep{Olson1999}.
Around $g=50\%$, $\mathcal{O}$ reaches $50\%$ and the dynamo is thus
of an $\alpha^2\Omega$-type. When increasing $g$ further the classical convective
columns practically vanish and the associated $\alpha$-effects decrease
significantly, leading to both weak poloidal and
toroidal fields (figure \ref{tor_pol_ome}, lower panel).
For the toroidal field the effect is somewhat compensated
by the growing $\Omega$-effect. The hemispherical dynamo clearly is
an $\alpha\Omega$-dynamo.

At $g=100\%$ the hemispherical mode clearly dominates and the
dynamo is of the $\alpha\Omega$-type with $\mathcal{O}\approx0.8$.
The Elsasser number has dropped to half its value at $g=0$ while
the magnetic Reynolds number has increased by a factor two (figure \ref{tor_pol_ome}, upper panel).
The hemispherical dynamo is clearly less effective than the
columnar dynamo. 

\begin{figure}[ht]
\vspace*{2mm}
\par
\begin{center}
\includegraphics[width=0.8\textwidth]{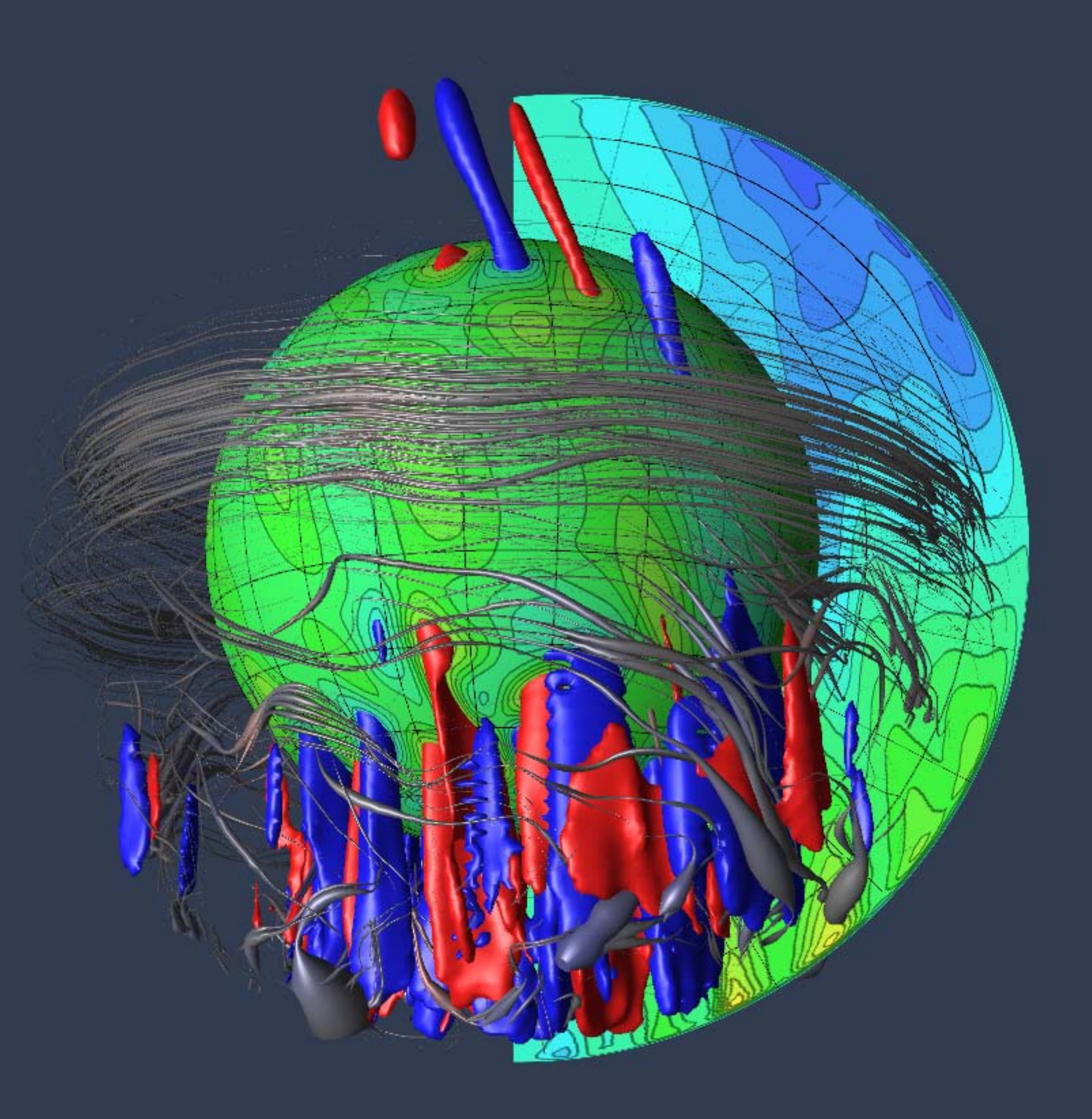}
\end{center}
\caption{3D visualization of flow and magnetic field generation in a
hemispherical dynamo. The meridional cut depicts contours of axisymmetric
zonal flows with prograde (retrograde) directions shown in yellow/red (blue).
Outward (inward) radial flows are shown as yellow/red (blue) contours of
a spherical shell at mid depth $r_{icb}+(r_{cmb}-r_{icb})/2$.
Red (blue) isosurfaces depict the 3D structure of convective upwellings
(downwellings). Gray fieldlines illustrate the magnetic field configuration.
Their thickness is scaled with the local magnetic field energy. See the article online-version for the color figure.}
\label{schema}
\end{figure}

Figure \ref{schema} illustrates the hemispherical dynamo mechanism in a 3D rendering.
Magnetic field lines show the magnetic field configuration, their
thickness is scaled with the local magnetic energy while red and blue colors
intensities indicate the relative inward and outward radial field
contribution. Plain gray lines are purely horizontal. Red and blue
isosurfaces characterize inward and outward
directed radial plume-like motions producing radial field magnetic field. Strong axisymmetric
zonal field is produced by a thermal wind related
$\Omega$-effect around the equatorial plane.

\subsection{Magnetic Oscillations}
\label{oscillations}
\begin{figure}[ht]
\vspace*{2mm}
\par
\begin{center}
\includegraphics[width=0.9\textwidth]{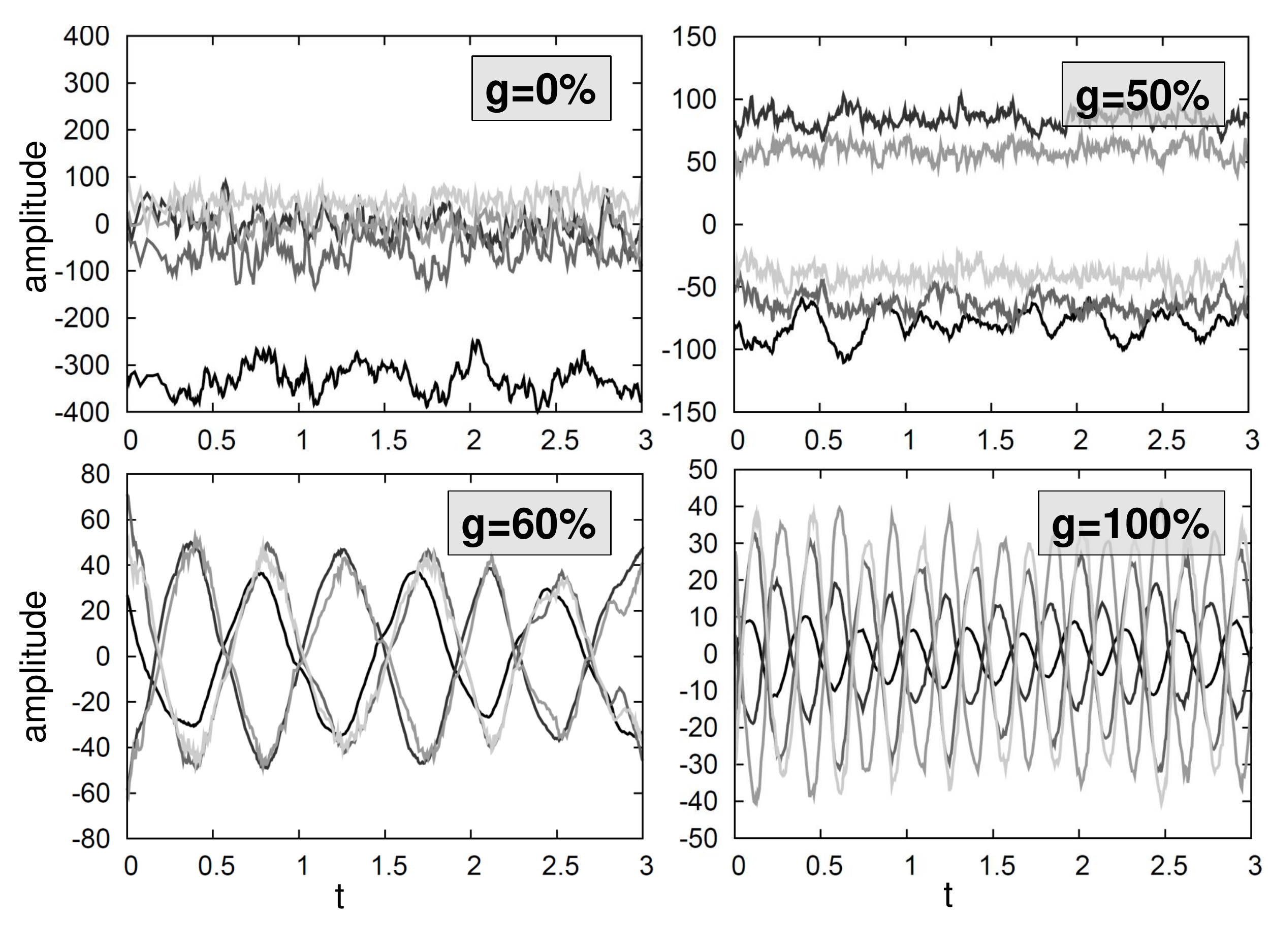}
\end{center}
\caption{Time evolution of the first five axisymmetric Gauss coefficients
at the CMB for the dipole dominated ($g=0$\%,$50$\%) (first and second panel),
the oscillatory ($g=60$\%, third panel) and the reversing hemispherical
regime ($g=100$\% bottom). The colours indicate different spherical
harmonic degrees $l$: black $l=1$, dark gray $l=2$, gray $l=3$, light gray $l=4$,
faint gray $l=5$. }
\label{gaussplots}
\end{figure}

Figure \ref{gaussplots} illustrates the changes in the time behavior of
the poloidal magnetic field when the CMB heat flux variation is increased.
We concentrate on axisymmetric Gaussian coefficients at the CMB $(r=r_{cmb})$
here. In the reference case $g=0$ (top panel) the axial dipole
dominates, varies chaotically in time and never reverses. If $g$ is increased
to $50\%$ (second panel) the relative importance of the axial quadrupole
component has increased significantly, which indicates the increasing
hemisphericity of the magnetic field. To yield a hemispherical magnetic field a similar amplitude in dipolar (equatorial antisymmetric) and quadrupolar (equatorial symmetric) dynamo family contributions is required \citep{Landeau2011, Grote2000}. 

When increasing the variation slightly to $g=60\%$ (third panel)
where the hemispherical mode finally dominates, all coefficients assume
a comparable amplitude and oscillate in phase around a zero mean
with a period of roughly half a magnetic diffusion time.
The faster convective flow variations can still be
discerned as a smaller amplitude superposition in figure \ref{gaussplots}.

\begin{figure}[ht]
\vspace*{2mm}
\par
\begin{center}
\includegraphics[width=1.0\textwidth]{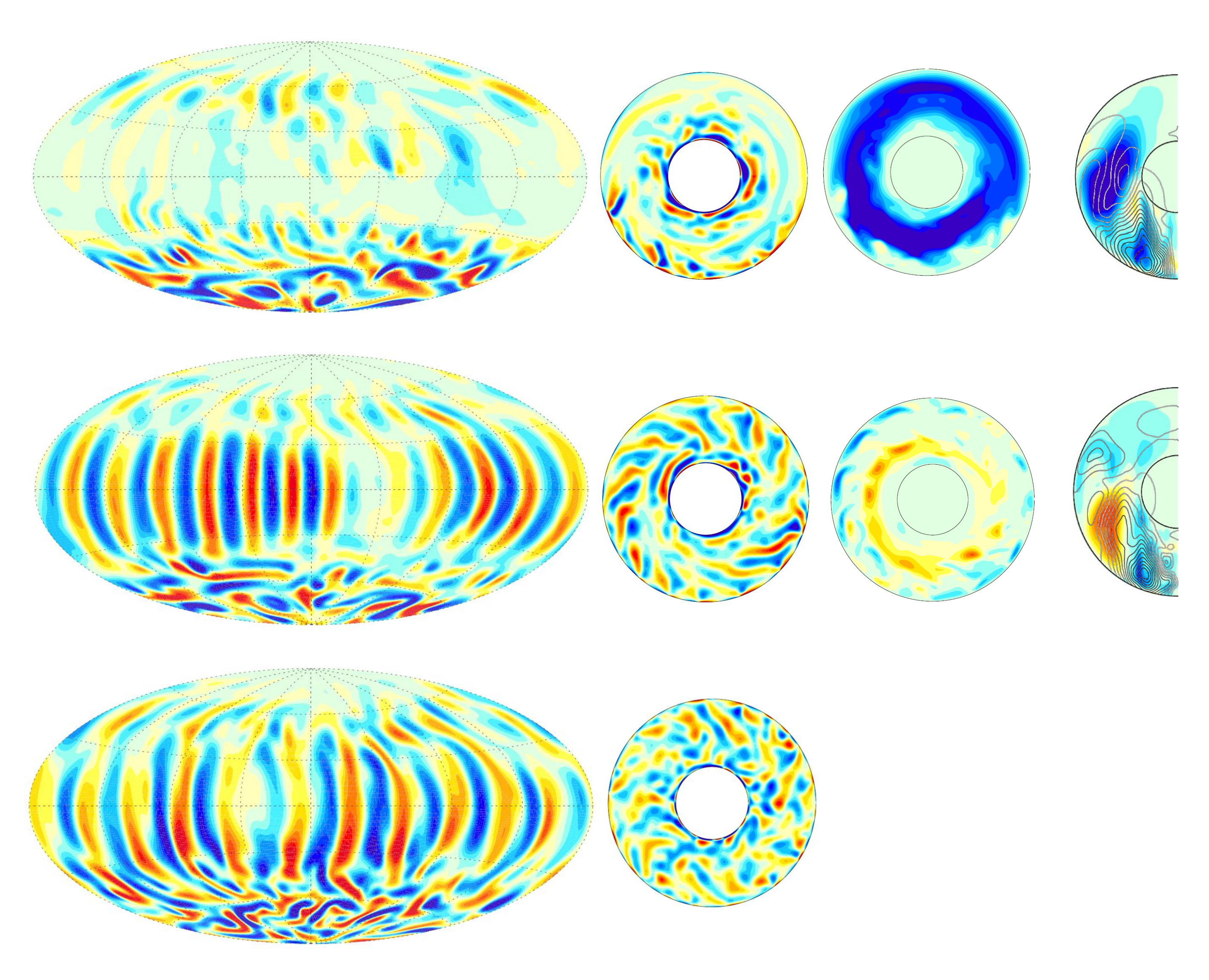}
\end{center}
\caption{The upper two rows depict the solution at $g=60$\% at
maximum ($\Lambda=12$) and minimum magnetic field amplitude ($\Lambda=0.06$) respectively. The lower
row shows a non-magnetic simulation at identical parameters for comparison.
Each row shows from left to right: the radial flow at mid depth in the shell,
the z-vorticity in the equatorial plane, the azimuthal magnetic field
at the equator and the zonal toroidal field. See the online-version of the article for the color figure.}
\label{lorentz}
\end{figure}

The oscillation is also present in a kinematic simulation
performed for comparison and is thus a purely magnetic phenomenon.
Lorentz forces nevertheless cause the flow to vary along with the
magnetic field. Since the coefficients vary in phase there are times
where the magnetic field and thus the
Lorentz forces are particularly weak or particularly strong.
Figure \ref{lorentz} illustrates the solutions at maximum (top) and minimum
(middle) rms field strength. At the minimum the convective columns are still
clearly present and the flow is similar to that found in the
non-magnetic simulations shown in the lower panel of figure \ref{lorentz}.
At the maximum the Lorentz forces, in particular those associated with the
strong zonal toroidal field, severely suppress the columns.
The magnetic field thereby further promotes the dominance
of the hemispherical mode \citep{Landeau2011}. This becomes even more apparent when comparing
the relative importance of the EAA mode $\mathcal{A}$ in magnetic and non-magnetic
simulations in the top panel of figure \ref{eaa_angle}. In the dynamo run $\mathcal{A}$ is
around $35\%$ higher than in the non-magnetic case for mild heat flux variation amplitudes.

When further increasing the amplitude of the CMB heat flux pattern, the frequency
grows, the time behavior becomes somewhat more complex, and
the different harmonics vary increasingly out of phase.
In addition, the relative importance of harmonics higher than the
dipole increase which indicates a concentration of the field
at higher southern latitudes.
The impact of the oscillations on the flows decreases since the
hemispherical mode now always clearly dominates and the relative
variation in the magnetic field amplitude becomes smaller.

The appearance of the oscillations may result from the increased
importance of the $\Omega$-effect which at $g=60$\% starts
to dominate toroidal field production (see figure \ref{tor_pol_ome}, lower panel).
The $\Omega$-effect could be responsible for the oscillatory behavior
of the solar dynamo as has, for example, been demonstrated by \citet{Parker1955}
who describes a simple purely magnetic wave phenomenon.
\citet{Busse2006} report Parker wave type oscillatory behavior in their
numerical dynamo simulations where the stress free mechanical boundary
conditions promote strong zonal flows and
thus a significant $\Omega$-effect.

\subsection{Arbitrary tilt angle}
\begin{figure}[ht]
\vspace*{2mm}
\par
\begin{center}
\includegraphics[width=0.8\textwidth]{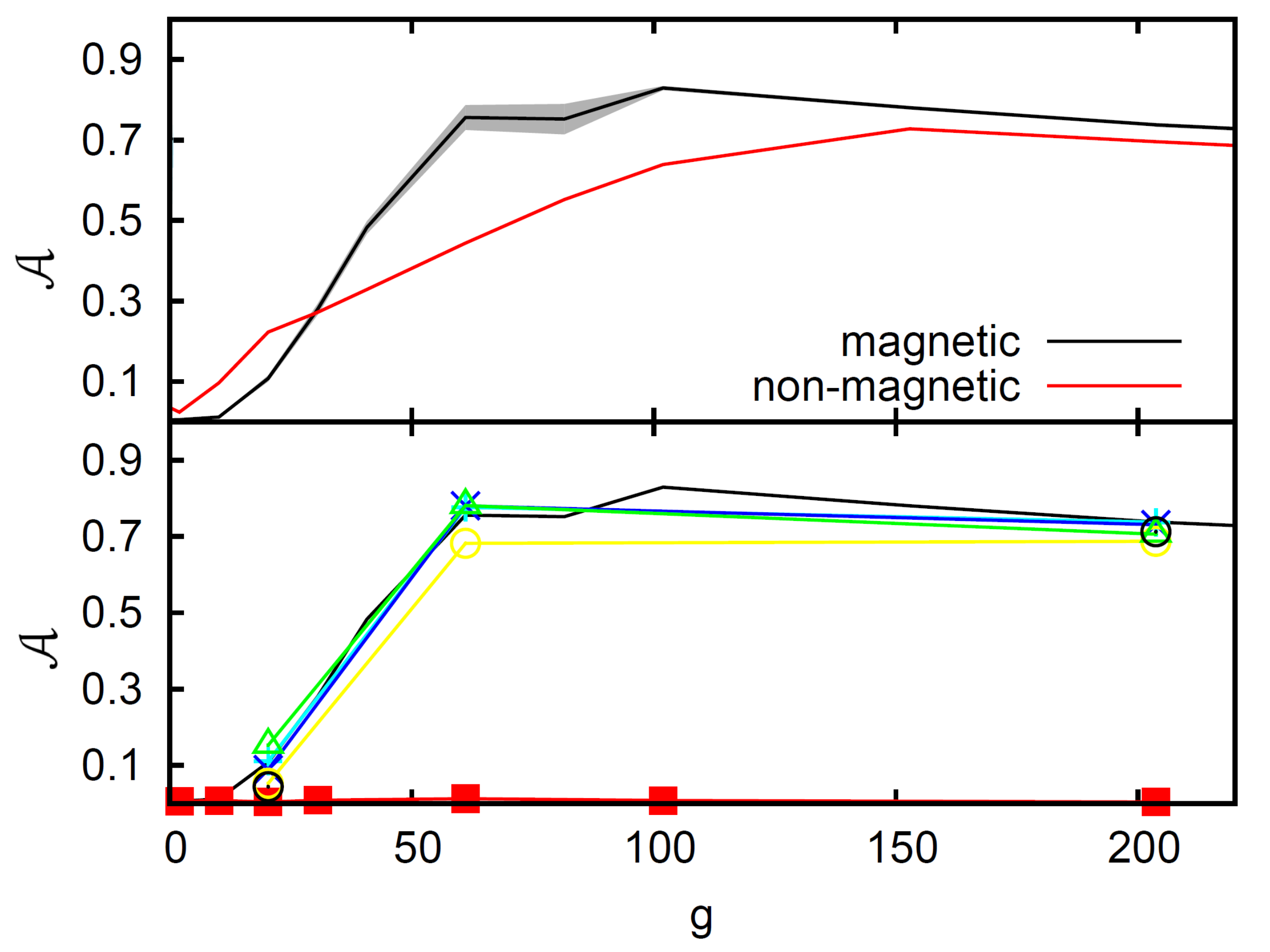}
\end{center}
\caption{upper panel : Effect of the Lorentz force on the relative kinetic energy
in the EAA mode for dynamo (black) and non-magnetic (gray) simulations.
The time variability is indicated by gray shaded areas in the width of
the standard deviation. The magnetic oscillation described in the text
lead to the stronger time variability in the dynamo simulations at
$g=60$\% and $g=100$\%. \newline
lower panel: The relative equatorially anti-symmetric and axisymmetric energy for
different tilting angles follows the onset of EAA convective mode in the
case for the axial pattern (black line). For the equatorial orientation (squares)
the EAA contribution to total kinetic energy remains Zero. Triangles - $10^\circ$,
crosses - $30^\circ$,  faint circles- $45^\circ$, dark circles - $60^\circ$, plus symbols -  $80^\circ$. See the online-version of the article for the color figure.  }
\label{eaa_angle}
\end{figure}

To explore to which degree the effects described above still hold when
the variation and rotation axis do not coincide we systematically
vary the variation pattern tilt angle $\alpha$ (see eq. \ref{defalpha})
up to 90 degrees.
The lower panel in figure \ref{eaa_angle} shows how $\mathcal{A}$, the relative
EAA kinetic energy, varies with
$g$ for different tilt angles.
Somewhat surprisingly, the hemispherical mode still clearly dominates for
tilt angles up to $\alpha=80^\circ$.
Only the rather special case of
$\alpha=90^\circ$ shows a new behavior, where $\mathcal{A}$ remains
negligible. It is thus the general breaking of the north/south symmetry that is essential
here. Since it leaves the northern hemisphere hotter than the southern it always leads to the above described dynamo mode.

\begin{figure}[ht]
\vspace*{2mm}
\par
\begin{center}
\includegraphics[width=0.7\textwidth]{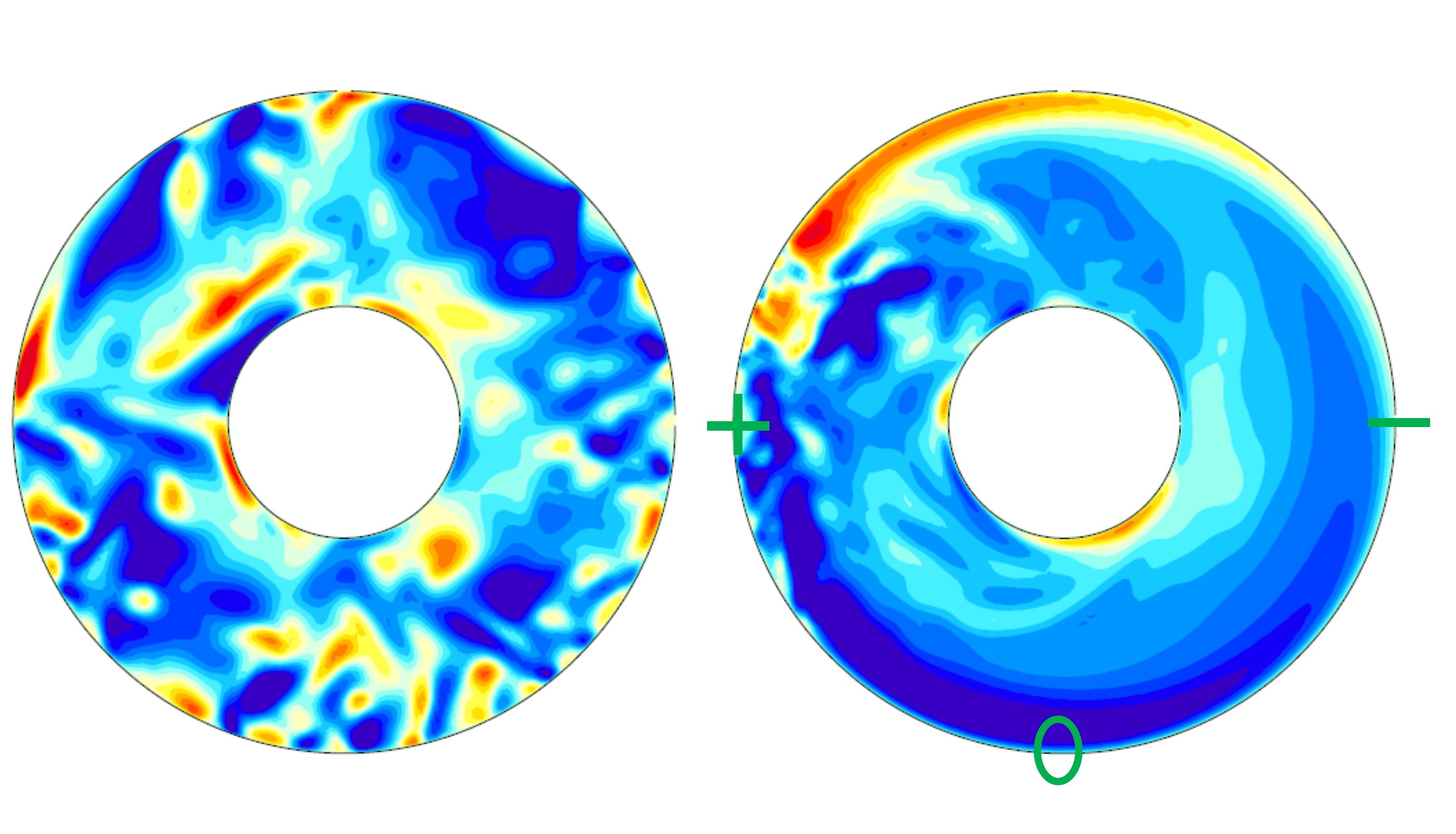}
\end{center}
\caption{Equatorial slice of $u_\phi$ for the homogeneous reference case (left)
and the equatorial heat flux anomaly (right). The plus, minus and zero character
describe the maximal, minimal and the zero line of the anomaly. See the online-version of the article for the color figure.}
\label{eq_vphi}
\end{figure}

\begin{figure}[ht]
\vspace*{2mm}
\par
\begin{center}
\includegraphics[width=0.7\textwidth]{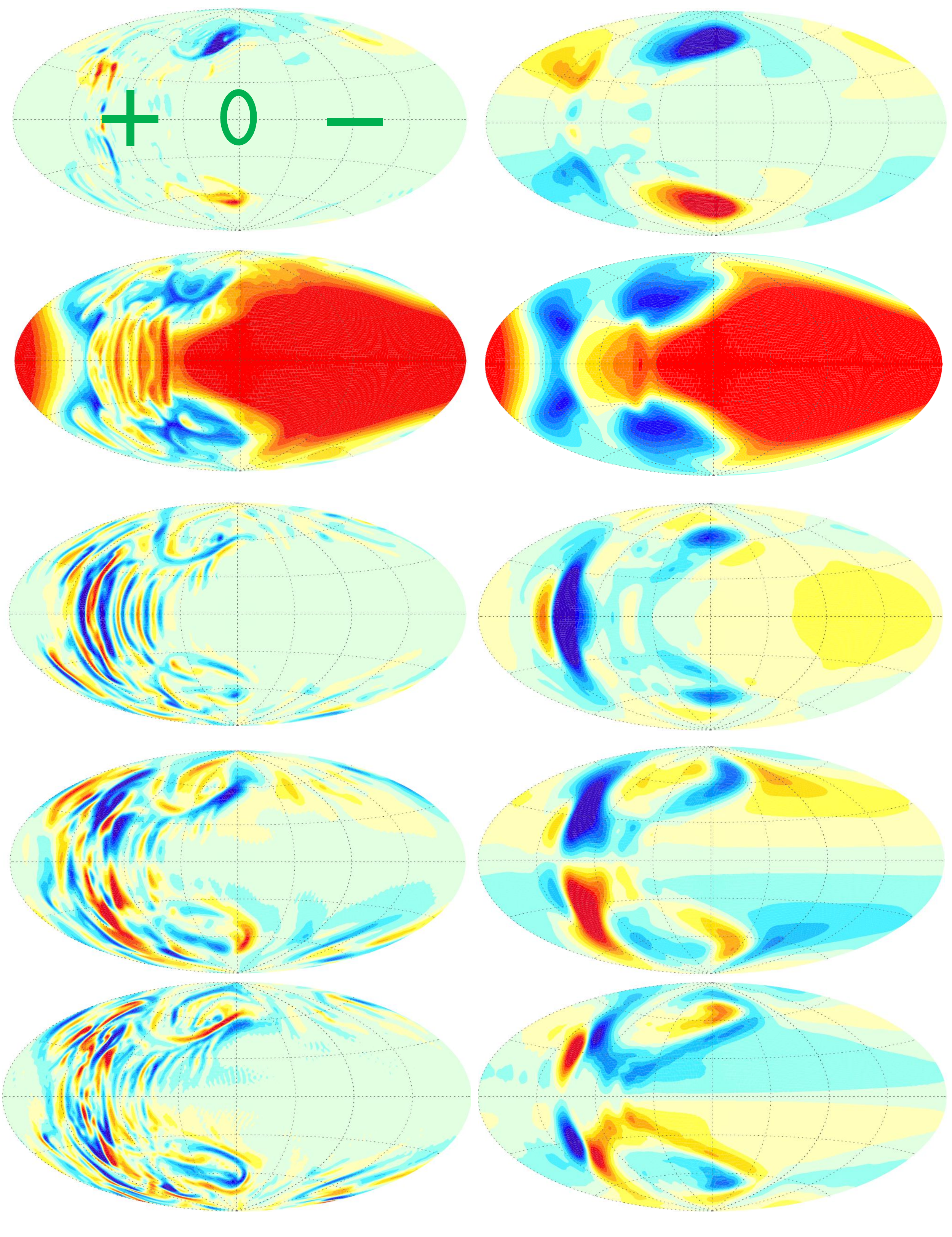}
\end{center}
\caption{Aitoff projections of spherical surfaces of (from top to bottom)
radial field, temperature (both at the CMB), $u_r$, $u_\theta$
and z-vorticity at $r/r_{cmb} = 0.8$ for a snapshot (left plots) and the
time average (right). Parameters: $Ra=4 \times 10^7$, $E=10^{-4}$, $Pm=2$, equatorial ($l=m=1$) perturbation with $g=100\%$ relative amplitude. See the online-version of the article for the color figure.}
\label{t90_flow}
\end{figure}

%Equatorial perturbation, convection and induction.
The $90$ degree tilt angle of the $(l=1,m=1)$ pattern
forms a special case because the breaking of the
north-south symmetry is missing here. Finally, the  effects of the
east/west symmetry breaking become apparent and
supersede the thermal wind related action in the other cases.
Figures \ref{eq_vphi} and \ref{t90_flow} illustrate the solution
for an equatorial anomaly with $g=200\%$.

 The resulting east/west temperature difference drives a large scale westward
directed flow and a more confined eastward flow in the equatorial region
of the outer part of the shell (figure \ref{eq_vphi}). Coriolis forces
divert the westward directed flow poleward and inward, and lead to the
confinement of the eastward directed flow.
Consequently, the westward flow plays the more important role here.

The diverted flows feed two distinct downwelling features that form
at the latitude of zero heat flux disturbance close to the tangent cylinder.
Due to the significant time dependence of the solution these can best be
identified in time average flows shown in figure \ref{t90_flow}.
Convective columns concentrated in the high heat flux hemisphere
but the center of their action is somewhat shifted retrograde, probably due
to the action of azimuthal winds. Other authors have shown that this
shift, for example, depends on the Ekman number \citep{Christensen2003}.
The  remaining columns are small scale and highly time dependent. On time average only one column-like feature remains, identified by a strong downwelling somewhat west to the longitude of highest heat flux. 

The time averaged flows form
two main vorticity structures illustrated in figure
\ref{t90_flow}. A long anticyclonic structure associated to the strong
equatorial westward flow stretches nearly around the globe and connects
the equator with high latitudes inside the tangent cylinder. A smaller
cyclonic feature is owed to the eastward equatorial flow.

The snapshot and time averaged radial magnetic fields shown in
figure \ref{t90_flow} are rather similar which demonstrates that
the time dependent small scale convective features are not very
efficient in  creating larger scale coherent magnetic field.
The radial field is strongly concentrated in patches above flow downwelling
where the associate inflows concentrate the background field \citep{Olson1999}.
Like in the study for dynamos with homogeneous CMB heat flux by \citet{Aubert2008}
the anti-cyclone mainly produces poloidal magnetic field.
The cyclone twists the field  in the other direction and therefore is responsible for
the pair of inverse (outward directed here) field patches located at
mid latitudes
in the western hemisphere. The exceptional strength of the
high latitude normal flux patches suggests that additional field line
stretching further intensified the field here. 

\subsection{Parameter Dependence}
\label{Parameter}
Focusing again at the axial heat flux anomaly we further study the influence of Rayleigh $Ra$, Ekman $E$ and magnetic Prandtl number $Pm$. In general we find that, independently of the Ekman $E$ and Rayleigh numbers $Ra$, a hemispherical dynamo mode is promoted once $g$ reaches a value of $60\%$. Close to the onset of dynamo action a mild variation can help to maintain dynamo action due to the additional $\Omega$-effect. See the cases at $E=10^{-4}$ and either $Ra=7\times10^6$, $Pm=2$ or $Ra=2\times10^8$, $Pm=1$ in table \ref{Tab1}. A strong amplitude of the heat flux anomaly can also suppress dynamo action due to the weakening of convective columns by the Lorentz force. For example, at $E=10^{-4}$, $Ra=4\times10^7$ and $Pm=2$ the dynamo fails once $g$ reaches $200\%$.

\begin{figure}[ht]
\vspace*{2mm}
\par
\begin{center}
\includegraphics[width=0.8\textwidth]{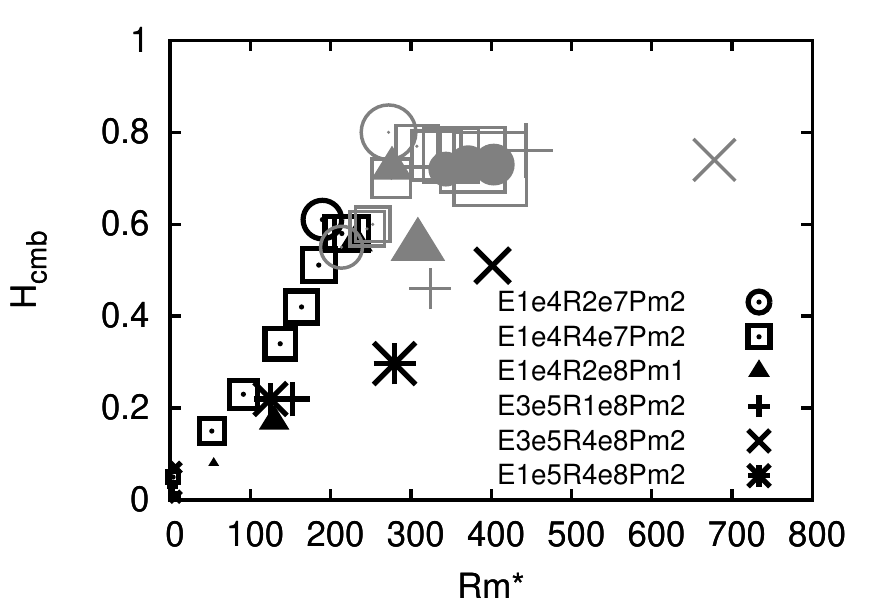}
\includegraphics[width=0.8\textwidth]{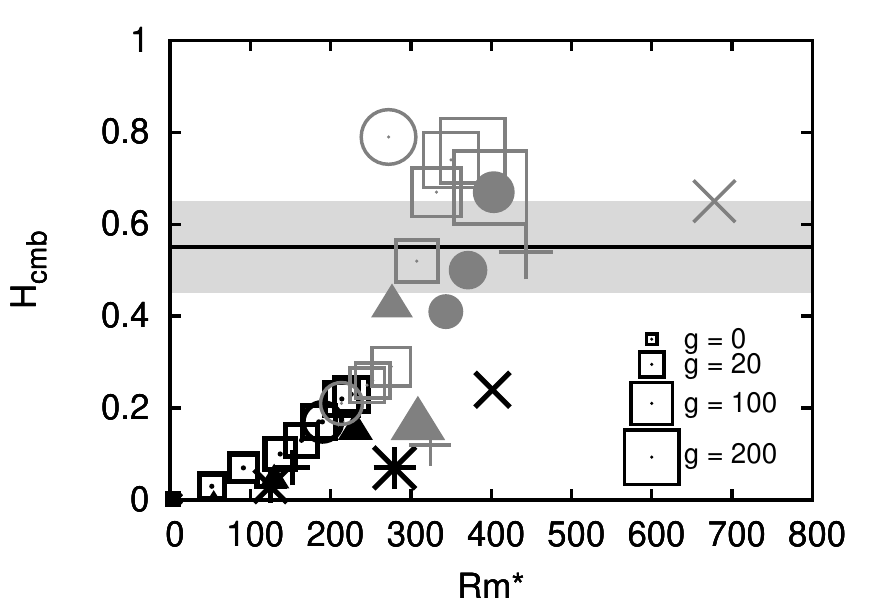}
\end{center}
\caption{Top panel: Hemisphericity at CMB versus \rmeaa, the magnetic Reynolds number
based on the equatorially anti-symmetric thermal wind. Oscillatory dynamos in gray, stationary in black symbols.\newline
bottom panel: Hemisphericity at the (imaginary) Martian surface
versus \rmeaa.}
\label{Hboth}
\end{figure}

Figure \ref{Hboth} shows how the CMB and surface hemisphericity (\hcmb\ , \hsur\ )  depends on the magnetic Reynolds number \rmeaa\ based on the equatorially anti-symmetric part
of the zonal flow only and therefore useful to quantify the important
$\Omega$-effect in the hemispherical dynamo cases.

For $E=10^{-4}$ the \hcmb\ values first increase linearly with \rmeaa\
and then saturates around $\hcmb\approx0.75$ for $\rmeaa\ge400$.
All cases roughly follow the same curve with the exception of the peculiar
$Ra=7\times10^6$, $Pm=2$ and $g=60$\% case described above.
This means that there is a trade off between $g$, $Ra$ and $Pm$; increasing
either parameter leads to larger \rmeaa\ values.
All the solution with hemisphericities $\hcmb\ > 0.6$ oscillate.

The few simulations at smaller Ekman numbers indicate that the degree of hemisphericity
decreases with decreasing $E$. This is to be expected since the Taylor Proudman theorem
becomes increasingly important \citep{Landeau2011}, inhibiting the ageostrophic hemispherical mode. Larger heat flux variation
amplitudes can help to counteract this effect. Since both inertia and
Lorentz forces can help to balance the Coriolis force, increasing
either $Ra$ or $Pm$ also helps.
For $E=3\tp{-5}$ an oscillatory case with $\hcmb=0.76$
is found for the larger Rayleigh number of $Ra=4\tp{8}$ and $g=100$\%.
For $E=10^{-5}$ $\hcmb$ remains small at $g=100$\% and we could
not afford to increase $Ra$ here since larger $Ra$ as well as lower $E$
values both promote smaller convective and magnetic length scales
and therefore require finer numerical grids.

The decrease in length scales has another interesting effect on the
radial dependence of hemisphericity. To yield a maximum hemisphericity,
equatorially symmetric ($l+m=$ even) and anti-symmetric ($l+m=$ odd) magnetic field contributions
must be of comparable strength, i.e.~obey a 'whitish' spectrum
(in a suitable normalization) \citep{Grote2000}. 
Since, however, the radial dependence of the modes depends on the spherical
harmonic degree (they decay like $r^{-(l+2)}$ away from the CMB) the
hemisphericity also depends on radius. The spectrum can only be perfectly
'white' at one radius. The smaller the scale of the magnetic field at
$r_{cmb}$ the further this radius lies beyond $r_{cmb}$.
This explains why the \hsur\ values shown in the lower panel
of figure \ref{Hboth} show a much larger scatter than the \hcmb\ values.
Larger values of $Ra$, $E$, but also $g$ and $Pm$ lead to
small magnetic scales and thus larger ratios of $\hsur$ over $\hcmb$.

\section{Application to Mars}
\label{applymars}
% I : rescaling, extrapolation, introduction to measurements
Could the hemispherical dynamo models presented above provide an explanation
for the crustal magnetization found on Mars? To address this question
we rely on the hemisphericity of the crustal magnetization
and the magnetic field strength inferred from Martian meteorites.
\citet{Amit2011} use MGS data to estimate a hemisphericity
between $\hsur=0.45$ and $\hsur=0.65$.
The magnetization of the Martian meteorite (ALH 84001) suggest a
field strength of the ancient dynamo between
$5$ and $50\, \textrm{$\mu$T} $ \citep{Weiss2002}.

Because our simulations show that the magnetic field
strength also varies significantly with the amplitude of the heat flux pattern, we rescale the dimensionless field strength in our simulations by
assuming that the Elsasser number provides a realistic value. Assuming a magnetic
diffusivity of $\lambda=1.32\,$m$^2\,$s$^{-1}$ and density of $\rho=7000\,$kg$\,$m$^{-3}$,
a rotation rate of $\Omega=7.1\times 10^{-5}\,$s$^{-1}$ and
the magnetic vacuum permeability then allows to rescale the Elsasser number
to dimensional field strengths. Time is rescaled via the magnetic diffusion
time $t_\lambda=D^2 \lambda^{-1}$ with an outer core radius of $1680\,$km.

\begin{figure}[ht]
\vspace*{2mm}
\par
\begin{center}
\includegraphics[width=0.7\textwidth]{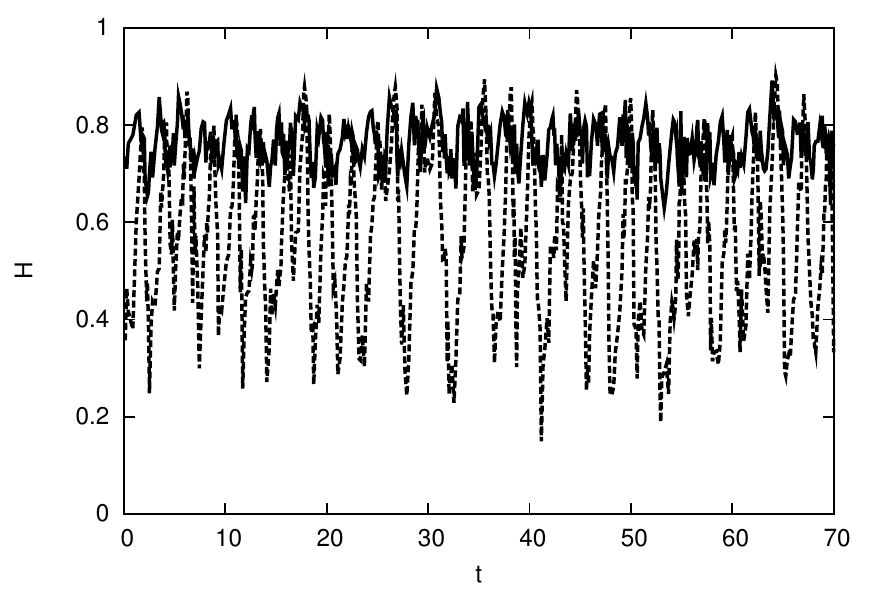}
\end{center}
\caption{Time evolution of hemisphericity $\mathcal{H}$ at the
CMB (solid) and surface (dashed) for $g=100\%$, $Ra=4 \times 10^7$, $E=10^{-4}$ and $Pm=2$.}
\label{H_rad_evo}
\end{figure}

% II : Hemisphericity, time evolution, radial dependence
We have included the Martian crustal hemisphericity values in the
lower panel of figure \ref{Hboth} to show that only oscillatory cases
fall in the required range with heat flux variation amplitudes
$g\ge 60 $\% and $\rmeaa\ge 300$.
Figure \ref{H_rad_evo} shows the temporal evolution of
\hcmb\ and \hsur\ for one of these cases.
The variation is surprisingly strong and oscillates at
twice the frequency of the individual Gauss coefficients.
Since all coefficients roughly oscillate with the same
period there are two instances during each period where the
hemisphericity is particularly large (around $\hemi\approx0.8$)
since axial dipole and quadrupole have the same
amplitude. Since the mean hemisphericity decreases with radius
the variation amplitude is much higher at the
planetary surface than at the CMB (figure \ref{H_rad_evo}).
The strong time dependence of oscillatory cases highlights that considerations
over which period the magnetization was acquired are extremely important.

To translate the dynamo field into a magnetization pattern, \citet{Amit2011}
suggest two end-members of how the magnetization was acquired. In the first end-member scenario called 'random' the crustal magnetization is acquired randomly in time and space and, according to \citet{Amit2011}, should reflect the time averaged intensity. In the second end-member called 'continuous', magnetization is acquired in global thick layers, so that the time intensity of the time average field is considered. However, since the magnetization records the magnetic changes happening during the slow crust formation, the local net magnetization, as seen by an observer, is always proportional to the time averaged local magnetic field possibly slightly dominated by the outermost layers. We therefore think, that the random magnetization scenario does not apply. The strong magnetization found on Mars indicates that a significant
portion of the crust is unidirectionally magnetized. \citet{Langlais2004} estimated a magnetization depth of $20 - 40\,$km depending on the magnetization density. Crust formation is a rather slow process that may take millions of years. Typical magnetic time scales can be much shorter. The  periods of the reversing strongly hemispherical dynamos discussed above, amount to not more than about ten thousand years.

Table \ref{Tab1} lists the time averaged rescaled magnetic field intensity at the model Martian surface. For $g\le 60$\% the field strengths are similar to that predicted for Mars \citep{Weiss2002} and fall somewhat below this values for larger $g$-values. In the strongly hemispherical oscillating cases, however, the amplitude of the time average field average to zero on time scales of the crustal magnetization. We therefore conclude that while the hemispherical dynamos can reach hemisphericities similar to that of the Martian crustal magnetization
their oscillatory nature makes them incompatible with the rather strong
magnetization amplitude.

\section{Discussion}
\label{Discussion}

We find that an equatorially anti-symmetric convective mode is consistently triggered
by a cosine heat flux variation that allows more
heat to escape through the southern than through the northern outer boundary of the dynamo region.
When the variation is strong enough, convective up- and down-wellings are
concentrated at the southern hemisphere and the northern hemisphere remains
hot. The associated latitudinal temperature gradients drive strong thermal
winds that dominate the flow when, for example, the variation amplitude $g$
exceeds $50\,$\% at $E=10^{4}$. Tilting the heat flux pattern axis leaves the solution more or less unchanged with the exception of the $90^\circ$-case where the equatorial symmetry remain unbroken. We conclude that breaking the equatorial symmetry is dynamically preferred over an equatorially oriented heat flux anomaly of the CMB heat flux. 

Due to the thermal winds, the dynamo type changes from $\alpha^2$ to
$\alpha\Omega$ but is generally less efficient. Lorentz forces
associated with the toroidal field created via the $\Omega$-effect tend
to kill whatever remains of classical columnar convection.
This further increases the equatorial anti-symmetry of the solution.
Poloidal fields are mainly produced by the southern up- and downwellings
which lead to a hemispherical field pattern at the outer boundary.

When the hemisphericity approaches values of that found in
Martian crustal magnetization, however, all dynamos start to oscillate
on (extrapolated) time scales of the order of $10\,$kyr.
These oscillations are reminiscent of previously described Parker waves in
dynamo simulations \citep{Busse2006}. As a typical characteristic of Parker waves, the frequency increases with the (square root of the) shear strength, see table \ref{Tab1}. The oscillation periods are much shorter than the time over which the deep reaching Martian
magnetization must have been acquired \citep{Langlais2004}.
Being a composite of many consecutive layers with alternating polarities
the net magnetization would scale with the time averaged field and
would therefore likely be much smaller than the predicted strength of the 
ancient Martian field magnetizing the crust \citep{Weiss2002}. The maximum hemisphericity for non-oscillatory dynamos amounts to a configuration where the mean northern field amplitude is only $50\%$ weaker than the southern. Additional effects like lava-overflows would then be required to explain the observed hemisphericity.

\citet{Amit2011, Stanley2008} also studied the effects of the identical sinusoidal boundary heat flux pattern and find very similar hemispherical solutions. \citet{Amit2011} used a very similar setup to ours and also reported oscillations when the dynamo becomes strongly hemispherical. \citet{Stanley2008} do not report the problematic oscillations intensively studied here, which may have to do with differences in the dynamo models. They study stress-free rather than rigid flow boundaries and assume that the growing inner core contributes to drive the dynamo while our model exclusively relies on internal heating. Should a hemispherical dynamo indeed be required to explain the observed magnetization dichotomy, this may indicate that ancient Mars already had an inner core. Alternatively efficient demagnetization mechanisms may have modified an originally more or less homogeneous magnetized crust \citep{Shahnas2007}. 

\citet{Landeau2011} observed that similar hemispherical
dynamos are found when the Rayleigh number exceeds a critical value. However, albeit the
effects are significantly smaller than when triggered via the boundary
heat flux.  All the cases explored here remain below this critical Rayleigh
number. \citet{Landeau2011} also mentioned that the equatorial anti-symmetry,
and thus the hemisphericity of the magnetic field, decreases when the
Ekman number is decreased. Our simulations at lower Ekman number seem to
confirm this trend although a meaningful extrapolation to the
Martian value of $E=3\times 10^{-15}$ would require further simulations
at lower Ekman numbers. To a certain extent the decrease can be
compensated by increasing the heat flux variation amplitude, the Rayleigh
number or the magnetic Prandtl number.

Our results show that a north-south symmetry breaking induced by lateral CMB heat flux variations can yield surprisingly strong effects. Fierce thermal winds and local southern upwellings take over from classical columnar convection and the dynamo changes from an $\alpha^2$ to an $\alpha \Omega$-type. The dominant $\Omega$-effect seems always linked to Parker-wave-like field oscillations typically discussed for stellar applications. It will be interesting to further explore the aspects independent of the application to Mars.

\section*{Acknowledgments}
The authors thank Ulrich Christensen for helpful discussions.
W. Dietrich acknowledges a PhD fellowship from the Helmholtz Research
Alliance 'Planetary Evolution and Life' and support from the 'International
Max Planck Research School on Physical Processes in the Solar System and Beyond'.

\bibliographystyle{plainnat}
\bibliography{mypaper}
\end{linenumbers}

\end{document}